\documentclass{article}
\usepackage{ieee_tns}
\usepackage[dvips]{epsfig}
\usepackage{cite}

\newcommand{\Fig}[1]{Fig.~\ref{#1}} 

\begin{document}

\title{The DISTO First Level Trigger at SATURNE}

\author{
     F. Balestra$^{d}$, Y. Bedfer$^{c}$, R. Bertini$^{c,d}$, L.C. Bland$^{b}$,
     A. Brenschede$^{h}$, F. Brochard$^{c}$, M.P. Bussa$^{d}$, 
     S. Choi$^{b}$, M. Debowski$^{e}$, M. Dzemidzic$^{b}$,
     J.Cl. Faivre$^{c}$, L. Fava$^{d}$, L. Ferrero$^{d}$, J. Foryciarz$^{g}$, 
     V. Frolov$^{a}$, R. Garfagnini$^{d}$, D. Gill$^{l}$, A. Grasso$^{d}$, 
     S. Heinz$^{c}$, W.W. Jacobs$^{b}$, W. K\"{u}hn$^{h}$,
     A. Maggiora$^{d}$, M. Maggiora$^{d}$, A. Manara$^{b,d}$, 
     D. Panzieri$^{d}$, H. Pfaff$^{h}$, 
     G.B. Pontecorvo$^{a}$, A. Popov$^{a}$, J. Ritman$^{h}$, P. Salabura$^{f}$,
     J. Stroth$^{i}$, V. Tchalyshev$^{a}$, F. Tosello$^{d}$, 
     S.E. Vigdor$^{b}$, G. Zosi$^{d}$   
      \\
      $^{a}$ JINR - Dubna \\
      $^{b}$ IUCF - Indiana \\
      $^{c}$ LNS - CEN Saclay \\
      $^{d}$ Dipartimento di Fisica``A. Avogadro'' 
                           and INFN - Torino \\
      $^{e}$ GSI - Darmstadt \\
      $^{f}$ Institute of Physics - Krakov \\
      $^{g}$ Jagellonian University - Krakow \\
      $^{h}$ II Physikalisches Institut - Giessen \\
      $^{i}$ Institut fur Kernphysik - Frankfurt \\
      $^{l}$ TRIUMF - Vancouver 
  }

\maketitle

\begin{abstract}
  The DISTO collaboration has built a large-acceptance magnetic spectrometer 
designed to provide broad kinematic coverage of multi-particle final states 
produced in $pp$ scattering.  The spectrometer has been installed in the 
polarized proton beam of the Saturne accelerator in Saclay to study 
polarization observables in the $\vec{p} p \rightarrow p K^{+} \vec{Y}$ ($Y = 
\Lambda, \Sigma^{0}$ or $Y^{*}$) reaction and vector meson production ($\phi, 
\omega$ and $\rho$) in $pp$ collisions.
  The common signature of such events is the multiplicity of four charged
particles in the final state.
  A flexible 1st level trigger which uses topological
information from fast detectors (scintillating fibers and hodoscope) has been 
built. 
  It is completely software programmable through a menu-driven user 
interface and allows switching between production
and monitor triggers on successive beam spills.
\end{abstract}


\section{Introduction}
\begin{figure}
   \centering\epsfig{figure=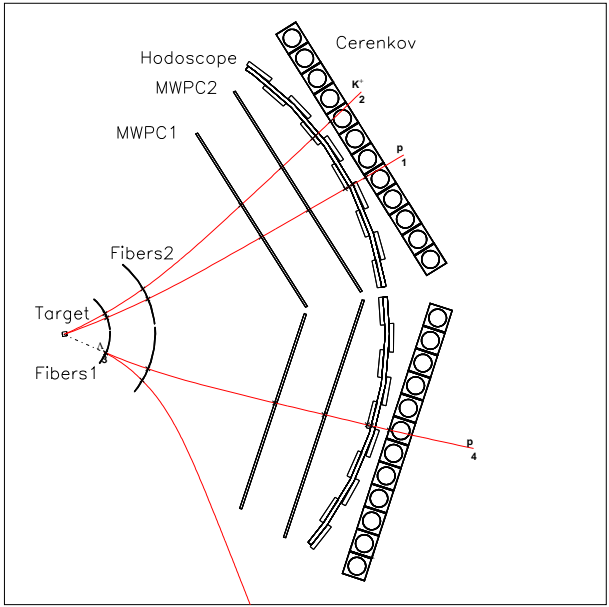,width=0.9\linewidth}
        \caption{\em Layout of the DISTO experimental setup, shown in plan 
                     view with a simulated $\Lambda$ event.}
        \label{fig_expapp}
\end{figure}

The DISTO collaboration has constructed a large-acceptance magnetic 
spectrometer to provide broad kinematic coverage of multi-particle final 
states involving charged particles produced in $pp$ scattering.  
Examples include the measurement of polarization observables in reactions such
as $\vec{p} p \rightarrow p K^{+} \vec{Y}$, with 
$Y$ representing a hyperon or a hyperon resonance ($\Lambda, \Sigma^{0}$ or 
$Y^{*}$), and the study of vector meson ($\phi, \omega$ and $\rho$) production 
in $\vec{p} p$ 
interactions~\cite{arvieux1,abegg,vigdor,arvieux2,bertini,maggiora}.
These measurements are now in progress at Laboratoire Nationale Saturne (LNS) 
in Saclay.  In experiment LNS-E213, the high-quality polarized proton beam, 
with kinetic energies ranging between 1.6 and 2.9 GeV, hits a liquid hydrogen 
target positioned in the center of a magnet that provides a strong magnetic 
field with cylindrical symmetry.  
The Saturne accelerator provides, at 2.9 GeV, every 4 s a spill of protons
with a flat top of 0.5 s. At 1.6 GeV the interspill time can be decreased down
to 2 s.
A sketch of the setup is shown in Fig.~\ref{fig_expapp}.  
Superimposed on the layout is a simulated event of the 
reaction $\vec{p} p \rightarrow p K^{+} \vec{\Lambda}$.

The detectors are arranged in two arms, to either side of a curved 
beam pipe (not shown in the figure). 
Tracking detectors comprise two pairs of cylindrical 
scintillating fiber chambers (two stereo layers, \textit{u-v} planes
at $+45^{\circ}$ and $-45^{\circ}$, and one 
\textit{y} plane with horizontal fibers), two pairs of \textit{u-v-x} planar 
Multi Wire Proportional Chambers (MWPC), a pair of $x-y$ cylindrical
scintillation counter hodoscopes, 
and a pair of planes of vertically segmented water Cerenkov counter
hodoscopes.  Light from the scintillating fibers is converted to 
electronic signals in a number of 80-anode Hamamatsu photomultiplier
tubes.

\vspace{4mm}

The 2492 fibers and 3202 wires of the tracking detectors are equipped with 
discriminator/latch electronics (LeCroy PCOS III).  Pulse height and timing 
information from the 32 scintillation counter
and the 48 Cerenkov counter photomultiplier tubes are 
read out via LeCroy FERA ADC's and TDC's.  Logic signals from
the scintillation hodoscope and the fiber chambers feed the level 1 trigger, 
which makes decisions based on the multiplicity of charged prongs and on the 
topology of the events.

%

\section{Overview of the 1st Level Trigger Architecture}

\begin{figure}
   \centering\epsfig{figure=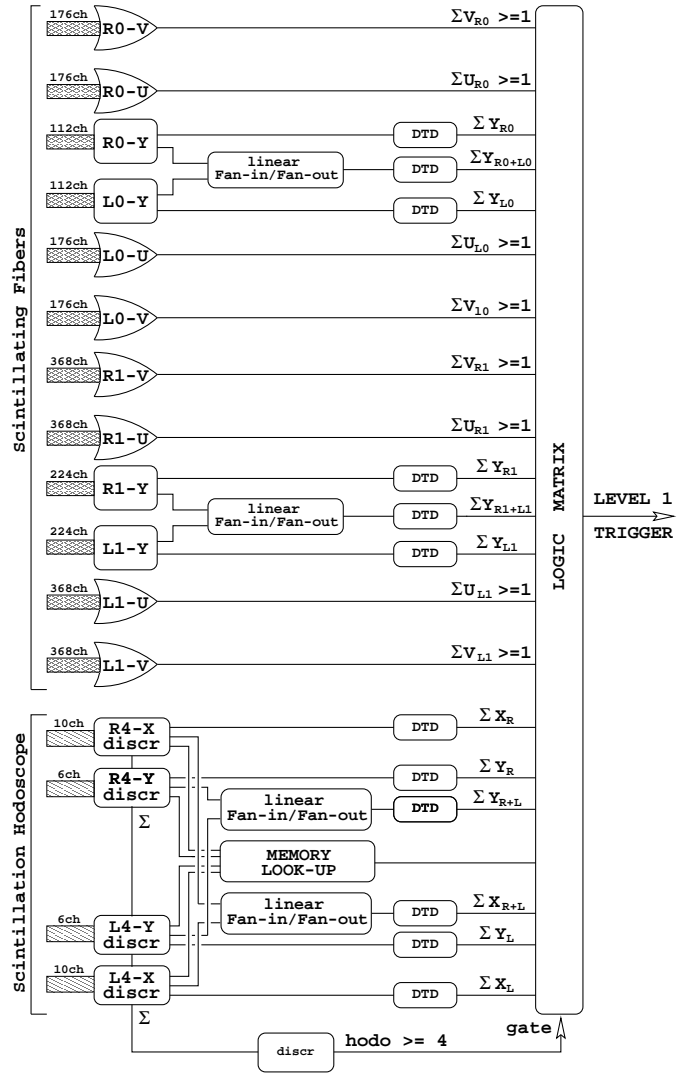,width=\linewidth}
\caption{\em Layout of the DISTO 1st level trigger.}
\label{fig_trig}
\end{figure}

The detailed scheme of the first level trigger is shown in \Fig{fig_trig}.
The front-end signals given by each detector are shown in the left 
part of the figure. 

\vspace{4mm}

Different processing techniques are applied to the raw signals given by 
the triggering detectors in order to match the information with the
first level trigger hardware setup.
The hodoscope analog signals (R4 and L4) are directly sent
to the counting room where they are processed by commercial programmable
CAMAC discriminator and delay modules.

The information from the fibers is processed near the detectors before being
sent to the counting room. The processing is different for the $y$ and for
the $u$ and $v$ planes.
The $y$ planes feed specially designed ``Current Adder'' NIM modules, 
described in detail in Section~\ref{sec:hardware}, that are able to ``spy''
the ECL bus signals and sink a shaped 2mA signal (i.e. 100mV on 50~$\Omega$
resistor) per hit fiber. These modules, together with commercial Linear 
Fan-in/Fan-out, generate a current signal proportional to the
input multiplicity.
The current signal given by the $y$ plane of each fiber chamber feeds a 
home-made ``Dual Threshold Discriminator'' (DTD) CAMAC module,
described in Section~\ref{sec:hardware}.
The module gives a ``true'' logic signal when the input falls
in a voltage window defined by two software progammable thresholds. 
Since the low and high thresholds are independently programmable, any 
$y$-plane multiplicity configuration can be selected.
 
In contrast, the signals from the $u$ and $v$ fiber planes are simply 
OR-ed using the internal wired-or facility of PCOS latches. 
The overall logical OR from each $u$ and $v$ plane is then sent to the
counting room, where the trigger can then demand only that at least
one hit be registered on the corresponding fiber plane.

The DTDs and a special version of Current-Adder modules
(without the internal shaper) are also used 
to get multiplicity information from different planes of the scintillation
hodoscope.

The multiplicity logic signals of the $y-u-v$ fiber planes and the hodoscope 
planes feed three 
commercial ``Logic Matrix Unit'' (LMU) modules where any logical combination 
of input signals (AND, OR, AND/OR mixed logic) can be programmed.

An additional circuit, not shown in \Fig{fig_trig}, synchronizes
the data acquisition with the time structure of the Saturne beam:
spills of 0.5 s duration separated by 3.5 s of acceleration overhead
at the maximum energy of 2.9 GeV.
In order to optimize the read-out throughput with minimal dead time,
the level-1 trigger directly starts the front-end readout sequence (with
PCOS first to allow for the slower FERA digitization) and the data 
are written, at 10MHz rate under a hard-wired arbitration system, into 
two daisy-chained, fast triple-port ECL memories (CES-8170).
The intervention of the data acquisition CPU is thus required only at the 
end of the Saturne beam spill to transfer the buffer stored in the memory to 
a free second level trigger CPU and thence to the recording device.
Using this operational mode we are able to reach a maximum acquisition
rate of more than 10000 events per spill. However, in actual running
conditions for the DISTO experiment, the multi-particle trigger rate
has been kept below 3500 events/spill to maintain FERA digitization
and readout dead times below 15\%.
These rates were obtained for an instantaneous luminosity of 
$10^{31}$ cm$^{-2}$ s$^{-1}$.

A special menu-driven interface, described in
Section~\ref{sec:soft_hum}, has been written to help the user program the 
trigger logic and multiplicity. 
The selected trigger choices are written into special files 
used by the data acquisition software to download into the trigger 
modules and copied into each event file stored on tape.

Taking advantage of the complete software programmability of the 
level-1 trigger system, it is possible in between successive beam
spills to
switch the experiment trigger from the main reaction to a monitor reaction
without introducing any additional dead time.
Indeed, we normally take data with a multi-particle trigger for 90\% of the 
Saturne spills, and with a different p-p elastic trigger for 
the remaining 10\% of the spills.  The p-p trigger is defined with the
aid of a commercial memory lookup module, used to identify the angular
correlation characteristic of elastic scattering kinematic
coincidences in the hit pattern of the scintillation hodoscope
elements.

\subsection{The 1st Level Trigger Hardware}
\label{sec:hardware}

Two special modules have been designed for the first level trigger of
the DISTO experiment. 

\begin{figure}
   \centering\epsfig{figure=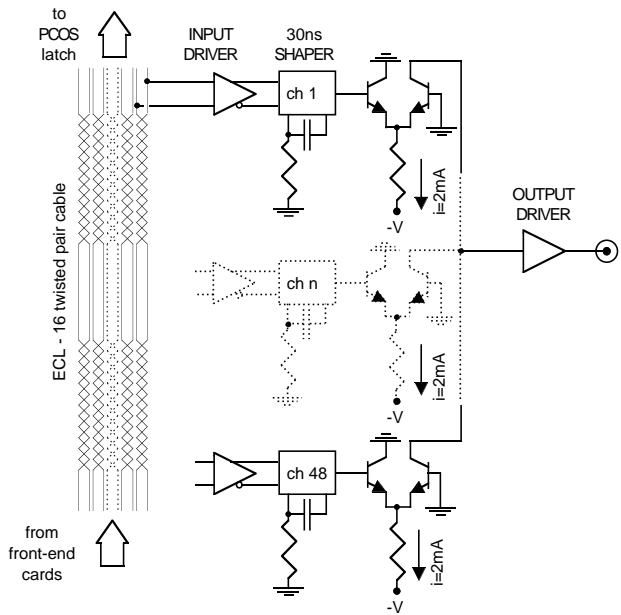,width=\linewidth}
\caption{\em Current Adder.}
\label{fig_curr_adder}
\end{figure}

The first one is the ``Current Adder'' shown in \Fig{fig_curr_adder}.
The main purpose of this module is to sample the ECL signals coming from
fiber chamber front-end cards, on their way to the PCOS latch modules.
The width of the input signal is internally shaped to 30ns to span
the time ranges associated with
intrinsic jitter of the photomultiplier signals and 
different lengths of delay cables and optical fibers. A high-impedance input 
driver is used in order to not disturb the proper termination of 
the signal at the PCOS input.
The output circuitry is able to sink and add a current of 2 mA per hit.
Up to 48 channels are housed in a single-width NIM module.

\vspace{4mm}

A special version of the module, without the internal shaping circuitry, has
been used for the hodoscope signals. The width of the 
signals in this case is previously adjusted by discriminator modules.
In all, twenty-two ``Current Adder'' modules have been built for the 
DISTO first level trigger. 

\begin{figure}
   \centering\epsfig{figure=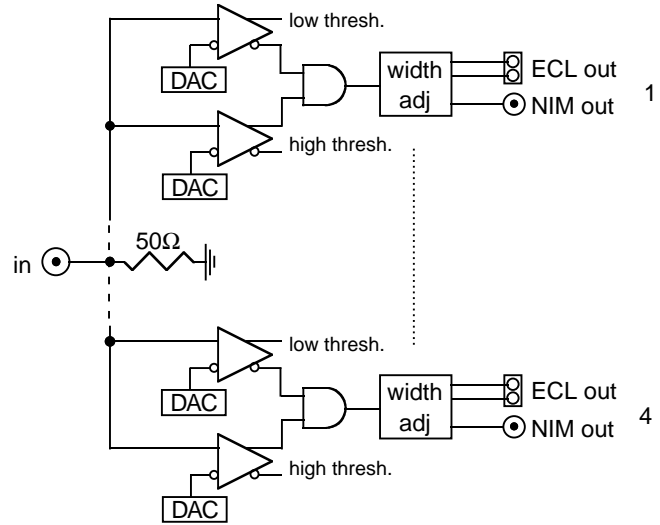,width=\linewidth}
\caption{\em Dual Threshold Discriminator (DTD).}
\label{fig_dual_thr}
\end{figure}

The second module specially designed for the DISTO first level trigger
is the ``Dual Threshold Discriminator'' shown in \Fig{fig_dual_thr}.

This module was designed to be jointly used with the ``Current Adder''. 
Nevertheless it is a fast, fully programmable discriminator, useful in
any experiment.
A single input feeds four independent internal sections.
Each section is composed of two fast comparators, one AND circuit and a
width adjustment.
The thresholds of the comparators can be set independently using 
standard CAMAC functions and can be read back for diagnostic purposes using
a multiplexed ADC (not shown in the figure).
Every section has two ECL outputs and one NIM output.
The input-output delay is 30ns.
Twelve DTD modules have been built and installed.

\vspace{4mm}

\subsection{The 1st Level Trigger Software and Menu-driven Interface}
\label{sec:soft_hum}

Interface software was written to help the experimenters to program
the first level trigger correctly. A minimum knowledge of the implemented
first level trigger hardware (\Fig{fig_trig}) is required.
For instance, the multiplicity request per detector is simply given as
the number of hits, instead of the digital value of the DAC's.
The internal consistency of the input data is checked and the 
program rejects inputs
when, for example, the value of the high threshold is below the low threshold 
or an unconnected input is selected.


\vspace{4mm}
\begin{figure}
   \centering\epsfig{figure=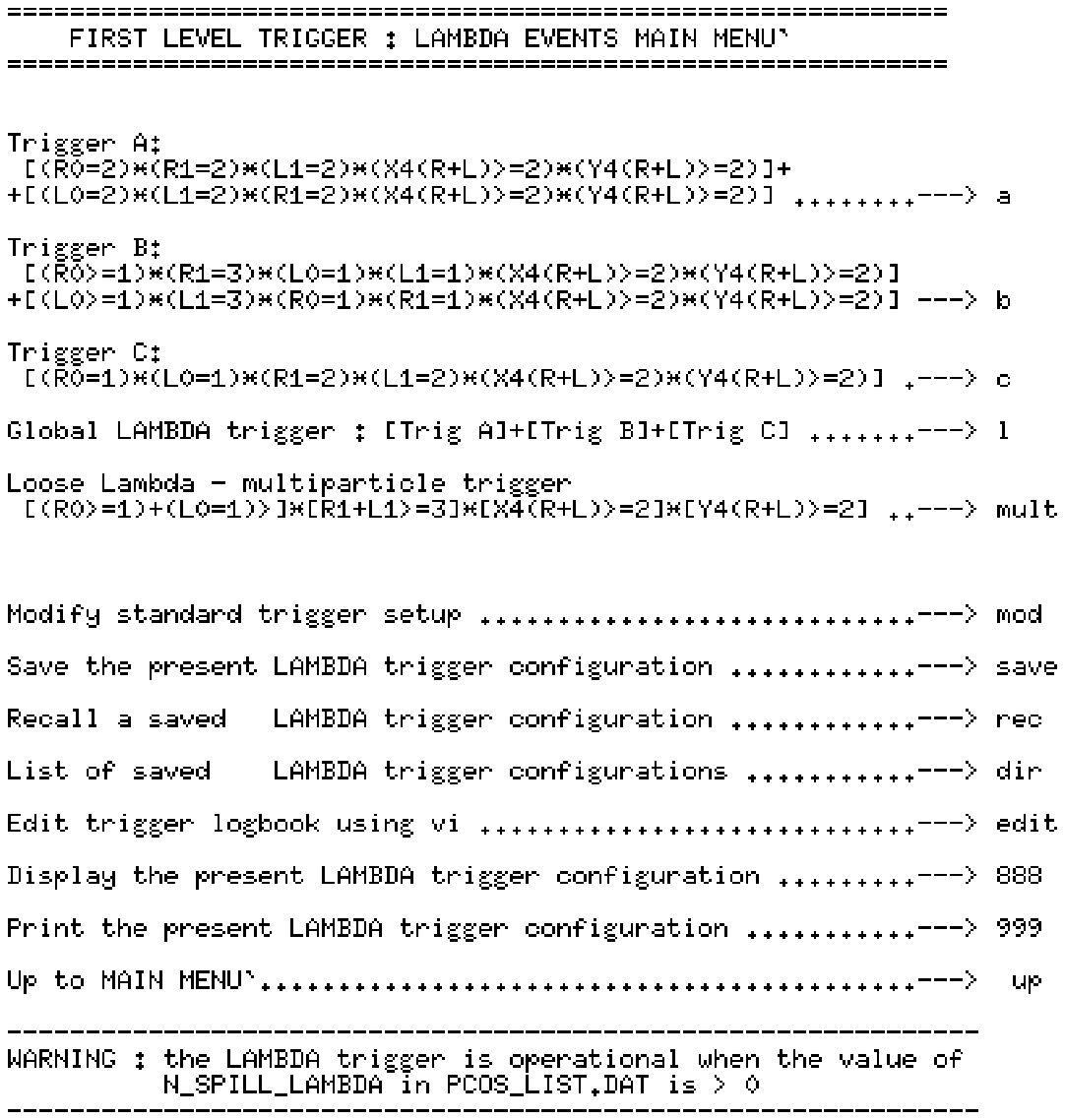,width=0.9\linewidth}
\caption{\em Example of the DISTO 1st level trigger men\'u.}
\label{fig_menu2}
\end{figure}

The human interface is menu-driven, with an example menu shown
in \Fig{fig_menu2}. 
The first menu allows selection of specialized sub-menus for the
$\Lambda$ (multi-particle) trigger, the p-p elastic monitor trigger, or 
diagnostic triggers used during detector tests.
The p-p monitor trigger sub-menu is similar to that for the $\Lambda$ trigger 
shown in \Fig{fig_menu2}.

The user can select some predefined trigger configurations, can load 
configurations previously saved, or can modify the present trigger setup.
In the last case the user enters into a series of sub-menus where any
configuration of DTD's or LMU's can be selected.
The trigger configuration can be saved and the trigger log-book (a special
file recording all modifications of the trigger) is updated.

\section{Conclusions}

A fully programmable trigger circuit has been built for the DISTO experiment 
at Saturne.
Thirty-four new modules specially designed and built 
for the experiment have been used together with commercial modules.
A menu-driven interface was designed to help in programming the trigger
hardware.

The programmability and flexibility of the trigger configuration allows us 
to monitor the experiment using a subset of p-p elastic events, comparing
monitoring data, taken under the same experimental conditions as the $\Lambda$ 
trigger data, with the vast and well known database available 
in the literature for p-p elastic scattering.
 
The trigger performance meets the design specifications for
selectivity and rate capability.

\section{Acknowledgments}

  We wish to thank S. Gallian and G. Maniscalco for their 
contribution to the data acquisition and detector electronics.  
  We are moreover grateful to G. Abbrugiati, G. Giraudo, M. Mucchi, and
M. Scalise for their efforts in the design and construction of scintillating 
fiber detectors.
  We are particularly indebted to N. Dibiase for essential
support since the beginning of the experiment. 
  The continous help of the whole Saturne staff was essential for the
success of this work.

\end{document}